\documentclass[pre,aps,twocolumn,showpacs]{revtex4}

\usepackage{graphicx}
\usepackage{bm}
\usepackage{amsmath}

\begin{document}

\title[Comb-drive MEMS Oscillators for Low Temperature Experiments]{Comb-drive MEMS Oscillators for Low Temperature Experiments}% Force 

\author{M. Gonz\'{a}lez}
% \\
\author{P. Zheng}%
\author{E. Garcell}%
\author{Y. Lee}%
\affiliation{Department of Physics, University of Florida, Gainesville, FL 32611, USA.}%

\author{H. B. Chan}

\affiliation{Department of Physics, The Hong Kong University of Science and Technology, Hong Kong, China.}%
\date{\today}

\begin{abstract}

We have designed and characterized micro-electro-mechanical systems (MEMS) for applications at low temperatures. The mechanical resonators were fabricated using a surface micromachining process. The devices consist of a pair of parallel plates with a well defined gap. The top plate can be actuated for shear motion relative to the bottom fixed plate through a set of comb-drive electrodes. Details on the operation and fabrication of the devices are discussed. The geometry was chosen to study the transport properties of the fluid entrained in the gap. An atomic force microscopy (AFM) study was performed in order to characterize the surface. A full characterization of their resonance properties in air and at room temperature was conducted as a function of pressure, from 10 mTorr to 760 Torr, ranging from a highly rarefied gas to a hydrodynamic regime. We demonstrate the operation of our resonator at low temperatures immersed in superfluid $^{4}$He and in the normal and superfluid states of $^{3}$He down to 0.3~mK. These MEMS oscillators show potential for use in a wide range of low temperature experiments, in particular, to probe novel phenomena in quantum fluids.

\end{abstract}



\maketitle

Mechanical oscillators such as torsional oscillators, vibrating wires, and quartz tuning forks have been used extensively at low temperatures for the study of quantum fluids. They allow the direct probe of properties of liquid helium and have provided irrefutable evidence of superfluidity \cite{andro}. These devices can be built in-house or are commercially available, providing a fair amount of flexibility in designing experiments. Modern silicon technology allows on-demand fabrication of micro-electro-mechanical systems (MEMS) with a plethora of functionalities as sensors and actuators. The capability to create devices and tailor their mechanical and electrical properties provides a rich arena in which novel experiments can be designed to explore new physics in quantum fluids at reduced length scales.

In recent years, micro/nano-mechanical structures have been incorporated in the low temperature study to reach the quantum mechanical ground state of mechanical resonators \cite{ADOConnell2010, TRocheleau2010} or to create devices similar to vibrating wires with the intention of using them for the study of liquid helium \cite{ECollin2008, ECollin2010, ECollin2011}. Predictions of a novel superfluid state in confined $^{3}$He \cite{ABVorontsov2007} have also stimulated the development of experimental probes capable of confining and probing the liquid's properties at the nanometer scale \cite{RGBennett2010, LVLevitin2010, SDimov2010}. Our motivation to develop the devices described in this paper stems from our interest in investigating novel phenomena occurring in superfluid $^{3}$He films. We envisioned a device that can form well-defined films of liquid and can simultaneously probe their properties with a high resolution. Our devices meet these design criteria. 

In this paper we describe the design and characterization of the MEMS oscillators. Details on the design, fabrication, and preparation are described in section I. The morphology of the surface is an important quality which could have an influence on the interaction of the device with its surrounding medium. We characterized the surface through atomic force microscopy (AFM) measurements. The results from these investigations are shown in section II. The general properties of the devices and the detection scheme are described in section III. The devices were studied at room and low temperatures. At room temperatures, the resonant properties were studied through a wide range of pressure, from 10~mTorr to 1~atm. Finally, we demonstrate the use of a device below 1~K while submerged in the superfluid state of liquid $^4$He and in the normal and superfluid states of liquid $^{3}$He down to 0.3~mK.

\section{\label{sec:level1}MEMS DEVICE FABRICATION AND SETUP}

\begin{figure}
\includegraphics[width=0.9\linewidth]{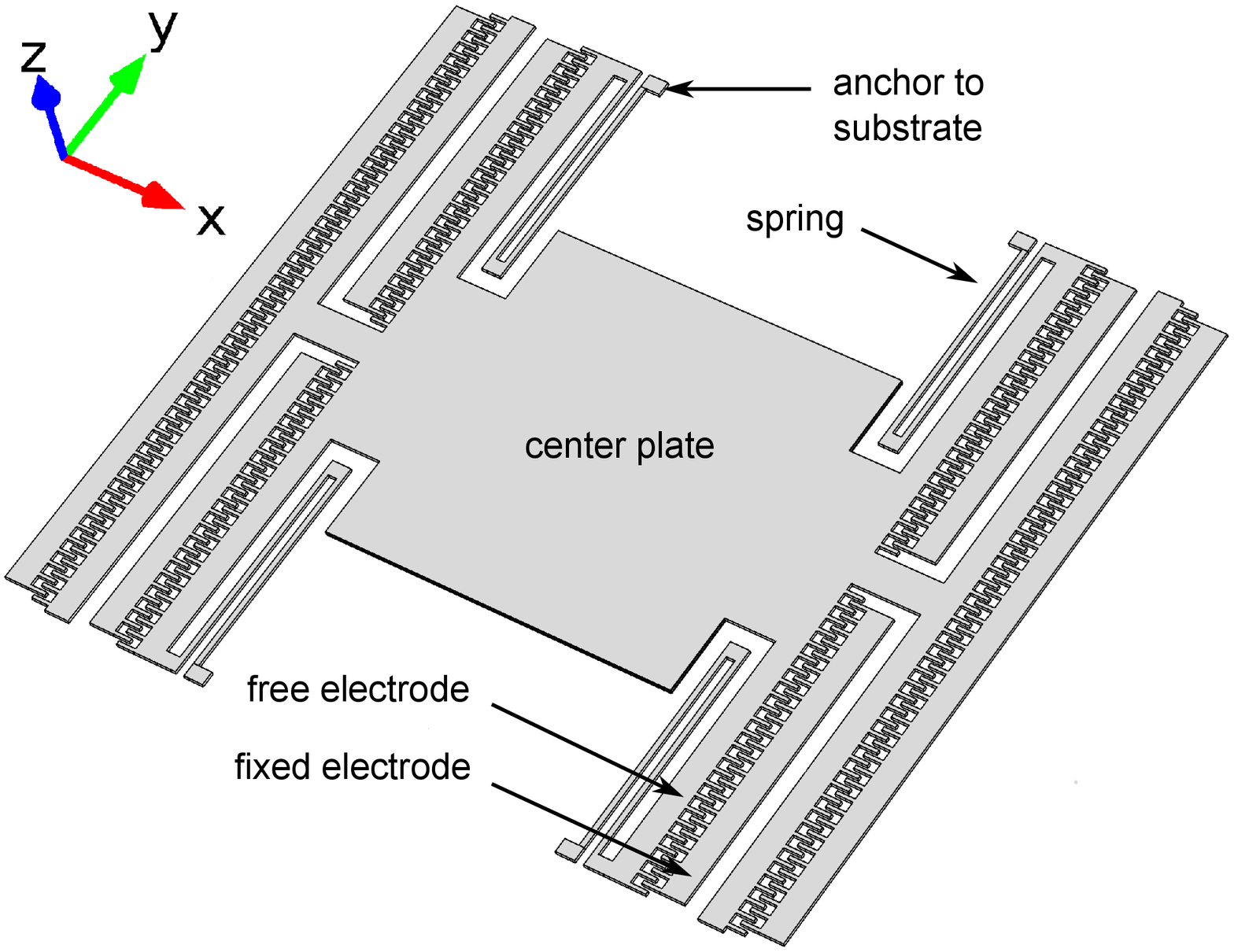} 
\caption{\label{fig:device}}{CAD image of a typical H1 device. No substrate is shown in this figure. The springs also serve as electrical connections from the center plate to the external bonding pads on the chip.}
\end{figure}

The geometry of the device is that of the so-called \emph{comb-drive} actuator. It is similar to the one used in modern MEMS accelerometers \cite{WCTang1989, WCTang1990}. The device consists of a movable center plate ($\approx$200$\times$200~$\mu$m$^{2}$) suspended by four serpentine springs. The center plate is maintained at a fixed distance above the substrate, thus creating a uniform gap. The two devices used throughout this work have the gaps of 1.25 and 0.75~$\mu$m, and are referred to as H1 and H2, respectively. A CAD image of a typical H1 device is shown in Fig.~\ref{fig:device}. The center plate can be set into lateral motion by electrostatic interaction between the inter-digitated comb-like electrodes attached to its sides and the fixed electrodes anchored to the substrate. For example, if a dc voltage is applied across one pair of electrodes, the center plate will be pulled towards the fixed electrode along the $x$-direction (see Fig.~\ref{fig:device}).

\begin{table}
\caption{\label{tab:table1} Thickness of the different layers used in PolyMUMPs}
\begin{ruledtabular}
\begin{tabular}{lcc}
Layer name   & Material        & Thickness ($\mu$m)  \\ 
\hline
Nitride      & Silicon nitride & 0.6    \\ 
Poly0        & Polysilicon     & 0.5    \\ 
First Oxide  & PSG             & 2      \\ 
Poly1        & Polysilicon     & 2      \\
Second Oxide & PSG             & 0.75   \\ 
Poly2        & Polysilicon     & 1.5    \\  
Metal        & Gold            & 0.75   \\ 
\end{tabular}
\end{ruledtabular}
\end{table}

We employed a shared wafer surface micromachining process called PolyMUMPs. The chips were fabricated by a commercial foundry called MEMSCAP. A layout of the different lithographic levels to be patterned on the chip was done using a specialized CAD software called L-Edit. The designs were then sent to the company for fabrication. In this process, the following layers are deposited on the substrate: three polysilicon layers (Poly0, Poly1, Poly2) used as the structural material, one silicon nitride layer (Nitride) used for electrical isolation from the substrate, two phosphosilicate glass (PSG) layers (First and Second Oxide) as sacrificial layers, and a metal layer (METAL) used for improved electrical connectivity. The label, material, and thickness of each layer patterned on the wafer is listed in Table~\ref{tab:table1}. The process is carried out on 100~mm wafers (1-2~$\Omega$-cm) and it starts by heavily doping the surface with phosphorus (9-10~$\Omega$/sq). The layers are deposited using low pressure chemical vapor deposition (LPCVD) and patterned by a combination of photolithography and reactive ion etching (RIE). PSG is used in the process to dope the different Poly layers with phosphorus by annealing the wafer at 1050$^\circ$C in argon during different steps of the process. The typical resistivity of a Poly 1 layer is $2.61\times 10^{-3}$~$\Omega$-cm. After depositing and patterning all the necessary layers to create the structures, the resulting chips are sent to the customer. More details can be found in the design handbook at the company's website \cite{memscap}.

\begin{figure}
\includegraphics[width=\linewidth]{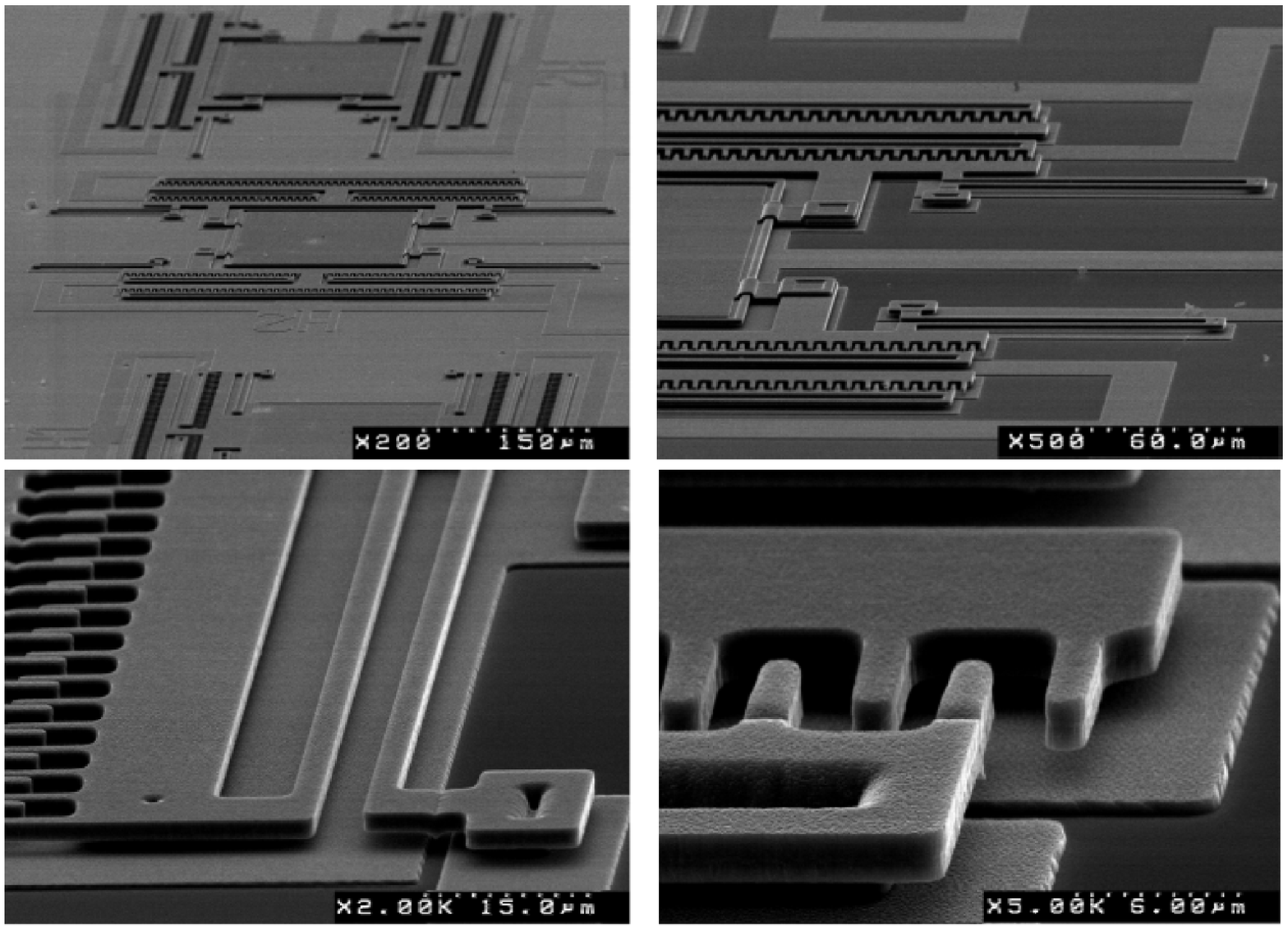} 
\caption{\label{fig:SEM}}{SEM pictures of a typical device. An overall view of an H2 device is shown in the top left figure. The top right figure shows half of the same device. The bottom left figure shows a more detailed view of the spring of an H1 device at its anchor point with the substrate. A close up look at the comb-electrode fingers is displayed in the bottom right figure.}
\end{figure}

The die received from the foundry was diced into smaller square chips of $\sim 2.5$~mm side length using a diamond saw. During this procedure, the devices were still coated by a photoresist layer, and the sacrificial layer was still present. This prevents damage to the device because the polysilicon structures are not yet movable. After dicing, the devices were ``released'' using a wet etching procedure. This removes the sacrificial PSG layers and frees the mechanically active parts. A standard process suggested by the foundry is followed for releasing \cite{memscap}. The etching is done by immersing the chips in 49\% hydrofluoric acid for 7 minutes. After the release process, if the chips are allowed to dry by exposure to the environment, the small plates comprising the devices might be pulled towards each other by capillary action and stick together. This phenomenon is called \emph{stiction}. To avoid this problem the chips were dried using a CO$_{2}$ critical point dryer. After drying, the chips were packaged in a custom designed 20-pin socket. Electrical connections were made from the bonding pads on the chip to the pads on the package by using a West-Bond ball wedge wire bonder with 0.025~mm thick gold wire. Scanning electron microscope (SEM) images of a typical released device are shown in Fig.~\ref{fig:SEM}.

\section{AFM Surface Characterization}
\begin{figure}
\includegraphics[width=0.9\linewidth]{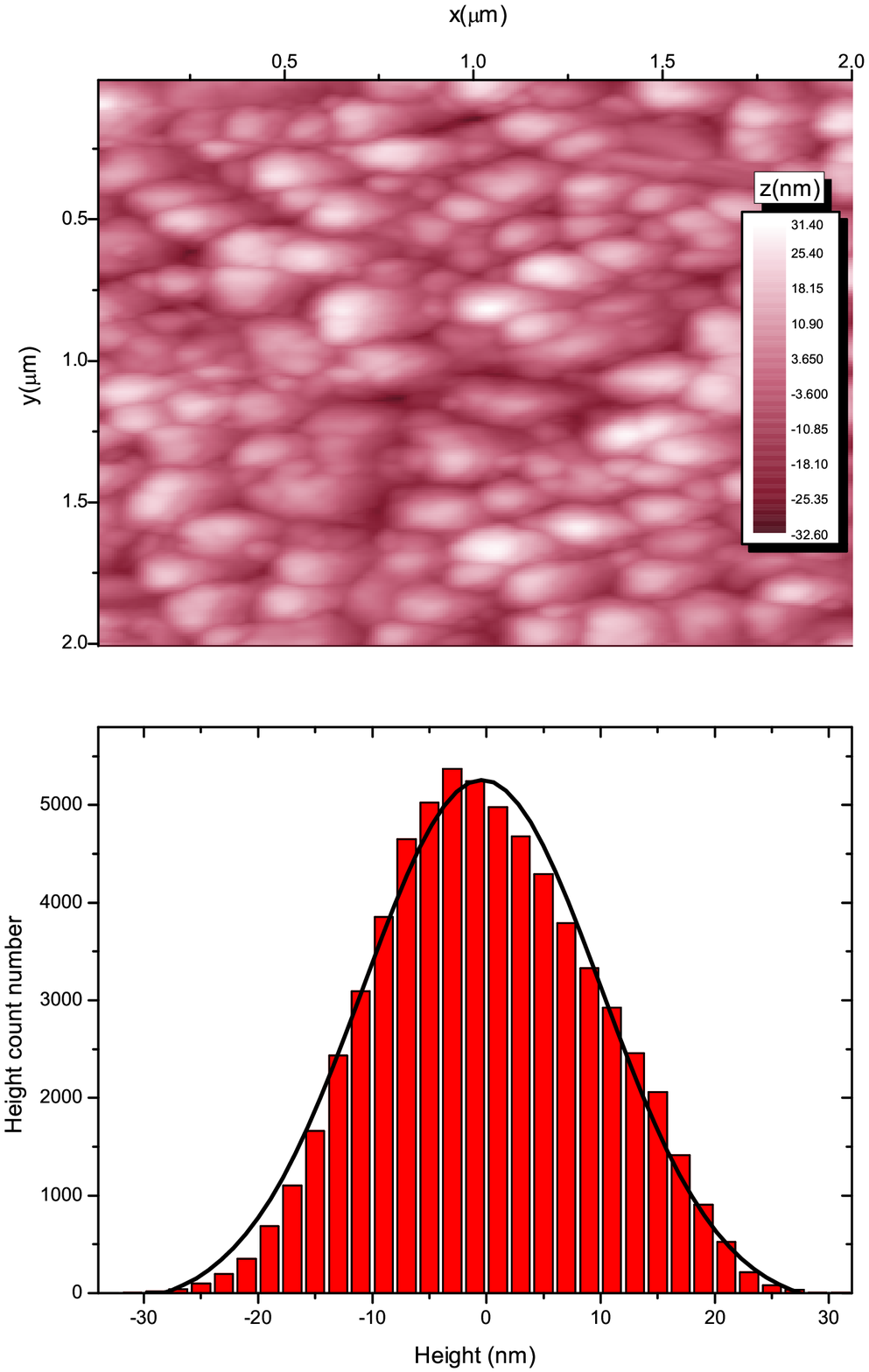}
\caption{\label{fig:AFM1}}{Top: Topography of the surface of the Poly 0 layer in a typical device. Bottom: Histogram of heights from the topography of the surface of a device. The solid line represents a fit to a Gaussian.}
\end{figure}
\begin{figure}
\includegraphics[width=0.9\linewidth]{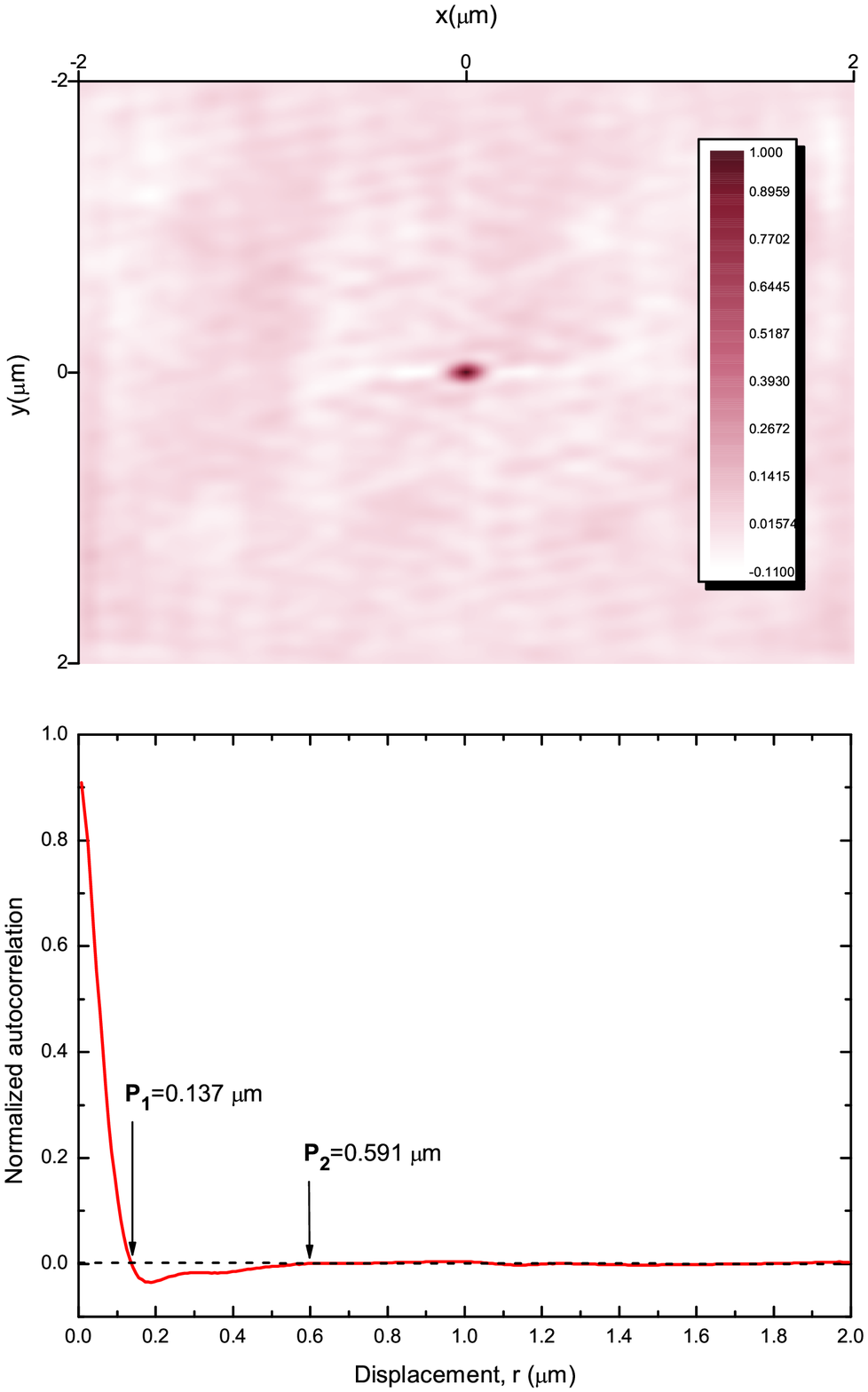}
\caption{\label{fig:AFM2}}{Top: Normalized autocorrelation function calculated from the topography of the Poly 0 surface in a typical device. Bottom: 1D normalized autocorrelation function averaged from different spots on the same device and from different devices contained in the same chip.}
\end{figure}
Experiments in thin films of $^{3}$He have found a possible non-trivial relation between the roughness of the surface bounding the liquid and the scattering relaxation time of quasiparticles close to the surface \cite{ACasey2004}. A theoretical description developed by Meyerovich \emph{et al.} transforms the problem of electrons propagating through a film with rough surfaces and a homogeneous bulk into that of a film with flat surfaces and a disordered bulk \cite{AEMeyerovich1998, AEMeyerovich2001, AEMeyerovich2002}. This theoretical scheme was adapted to explain recent experiments in which an anomalous temperature dependence was observed in the transport properties of a thin film of liquid helium \cite{PSharma2011}. This theory uses, as an input, information about the surface's autocorrelation function, which is related to the momentum transition probability of quasiparticles scattering off the surface. It is, therefore, imperative to have a detailed understanding of the topography of the polycrystalline plates in our devices. For this purpose, we have carried out AFM measurements on a sample chip containing the MEMS resonators.

The chip was initially sprayed with nitrogen gas to blow off the top plate of all the devices present on the chip. A Nanoscope IIIA Scanning Probe Microscope was used to scan the exposed Poly 0 or Poly 1 surfaces that constitute the bottom plate of a typical device. Twelve different spots were probed with scans of $10\times 10$, $2\times 2$, and $0.4\times 0.4$~$\mu$m$^{2}$ on different parts of the same device and on different devices contained in the same chip. The AFM tip was set in tapping mode with a resonance frequency of 69.16~kHz. Data for a scan size of $2\times 2$ ~$\mu$m$^{2}$ are shown in Fig.~\ref{fig:AFM1}. The data was flattened and plane fitted. A histogram of heights shows a symmetric Gaussian distribution about the mean plane, which is indicated as zero height. From the full width at half maximum (FWHM) of the Gaussian, the average feature height with respect to the mean plane is estimated to be $10.4$~nm. 

\begin{table*}
\caption{\label{tab:table2}Calculated device properties from layout geometry}
\begin{ruledtabular}
\begin{tabular}{lccccc}
Device & Gap, $d$ &  Area, $A_{s}$              & Mass, $m$           & Spring, $k_{s}$          & Capacitance, $c_{0}$        \\ 
\ & ($\mu$m) & ($\mu$m$\times\mu$m)  & ($\times 10^{-10}$~kg)  & (N/m)          & ($\times$10$^{-3}$~pF)      \\ 
\hline
H1$_{t,m}$   &1.25 &$192\times192$ &3.45 &2.47 &8.41  \\ 
H2   &0.75    &$178\times 178$ &2.77 &2.47 &8.41  \\
\end{tabular}
\end{ruledtabular}
\end{table*}

\begin{figure*}
\includegraphics[width=0.32\textwidth, angle=-90, trim=4.7cm 6.1cm 9.9cm 9.4cm, clip=true]{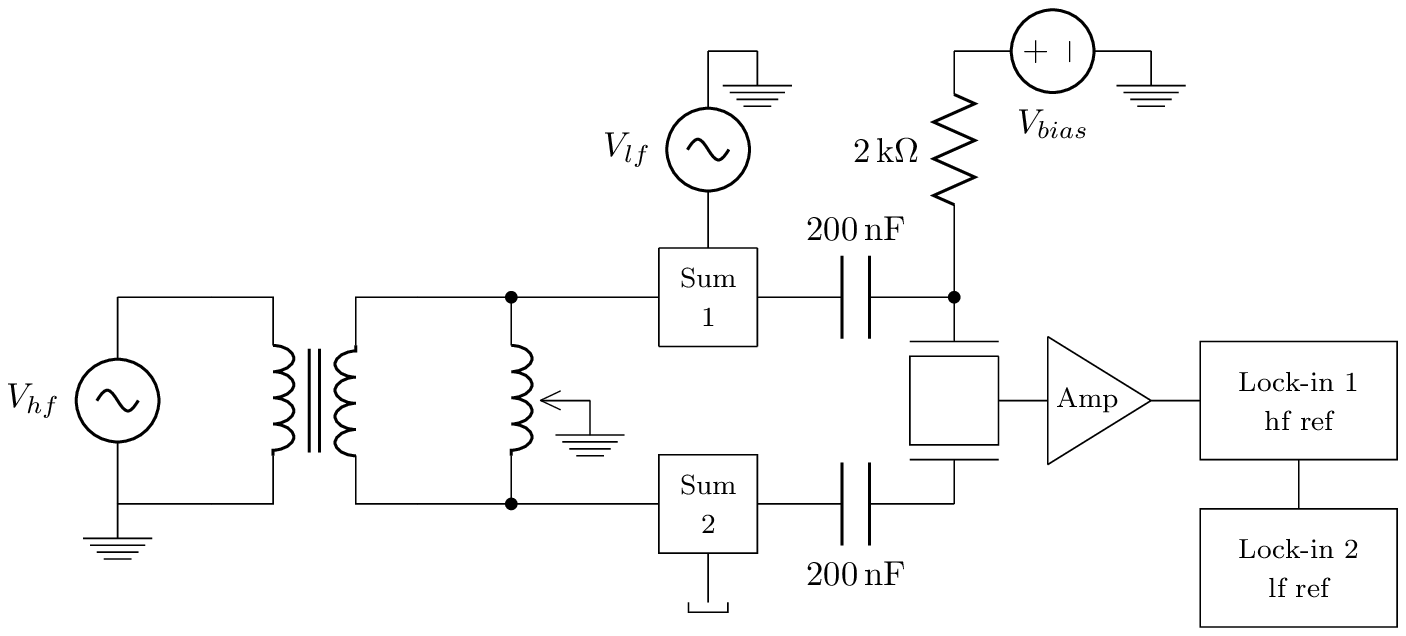}
\caption{\label{fig:circuit}}{Circuit diagram for the capacitance bridge technique employed. An external ratio transformer is used as a tunable inductive voltage divider along with the two capacitors formed by the two sets of electrodes at the sides of the device.}
\end{figure*}

A normalized autocorrelation function was calculated from the surface topographies to investigate the correlation between grains. As shown in Fig.~\ref{fig:AFM2}, the function is sharply peaked at zero displacement. From the autocorrelation corresponding to the $2\times 2$ ~$\mu$m$^{2}$ topography, a one-dimensional autocorrelation was calculated by averaging all the values on the perimeter of a circle of radius $r$ centered at the origin of the contour plot (top panel of Fig.~\ref{fig:AFM2}). The resulting autocorrelation function is shown at the bottom of Fig.~\ref{fig:AFM2}. The autocorrelation goes from a positive to a negative value at point $P_{1}$, where $r=0.137$~$\mu$m. This distance provides an estimate of the average grain lateral size. At the second point, $P_{2}$, the autocorrelation goes back to a positive value. This point indicates the existence of another feature on the surface of average size $0.591$~$\mu$m. Clusters of grains of approximately this size can be seen in the topography of the surface. In summary, from the height histogram, grains can be estimated to have a height of $10.4$~nm. From the autocorrelation, the estimated width is $0.137$~$\mu$m. Furthermore, an average cluster size can be estimated to be $0.591$~$\mu$m from the autocorrelation function. The last two distances establish two length scales: a short length scale at 137~nm and a long length scale at 591~nm. These can be verified by visual inspection of the topography (see Fig.~\ref{fig:AFM1}).

\section{Device Properties and Characterization}

\begin{table*}
\caption{\label{tab:table3}Resonance frequencies and measured Q-factors in high vacuum ($\sim 3$~mTorr)}
\begin{ruledtabular}
\begin{tabular}{lcccc}
Device & $f_{c}$ (calculated) & $f_{s}$ (simulated) & $f_{m}$ (measured) & $Q-factor$ (measured)  \\ 
\      & (Hz)                 & (Hz)                & (Hz)               &                        \\ 
\hline
H1   &29133.3 &24755.4 &23063.1 &6424.3    \\ 
H1$_{t}$ &29133.3 &24755.4 &22006.2 &23188.8   \\ 
H1$_{m}$ &29133.3 &24755.4 &21516.4 &105472.6  \\
H2   &32513.2 &28438.3 &26691.6 &5858.6    \\
\end{tabular}
\end{ruledtabular}
\end{table*}

Some important device specifications for both types of devices, H1 and H2, are shown in Table~\ref{tab:table2}. These include: the gap distance $d$, the area of the main plate $A_{s}$, the total mass of the movable plate $m$, the single spring constant $k_{s}$, and the capacitance $c_{0}$. The total spring constant is $4k_{s}$. The total capacitance, $c_{0}$, between a fixed electrode and its movable counterpart can be calculated from the expression \cite{MBao2005}:
\begin{equation}
\label{eq:capacitance}
c_{0}=n\frac{A\epsilon\epsilon_{0}}{d_{f}},
\end{equation}
where $A$ is the total overlap area, $d_{f}$ is the distance between fingers, $n$ is the total number of fingers, $\epsilon_{0}$ is the permittivity of free space, and $\epsilon$ is the relative permittivity. 

The spring constant is determined by the Young's modulus of polysilicon and the geometry of the spring. For a serpentine spring with $N$ guided beams ($N=2$ in our devices), the spring constant is given by \cite{VKaayakari2009}
\begin{equation}
k_{s}=E\left(\frac{w}{l}\right)^{3}\frac{\tau}{N},
\end{equation}
where $E$ is the Young's modulus, $w$ is the width, $\tau$ is the height, and $l$ is the length of the spring beam. The lithographic masks used in the process by MEMSCAP have a pixelation of $0.25$~$\mu$m \cite{memscap}. If an error as high as $0.5$~$\mu$m per lateral dimension is assumed, the total error in the spring constant can be as high as $38$\%. The mass of the device was calculated directly from the geometry of the devices. In this case the error in the lateral dimensions is negligible, and the only relevant source of the error is the thickness of the Poly layers (within $\sim 10$~nm) \cite{memscap}. This introduces an error of less than $1$\% in the estimation of the mass. The resonance frequencies obtained from the values of the calculated mass and the spring constant are listed in Table~\ref{tab:table3}. We used $w=3$~$\mu$m, $\tau=2$~$\mu$m, and $l=120$~$\mu$m. A Young's modulus value of $E=158\pm 10$~GPa was used (see Ref.~\cite{memscap}). 

\subsection{Detection Scheme}
\begin{figure}
\includegraphics[width=\linewidth]{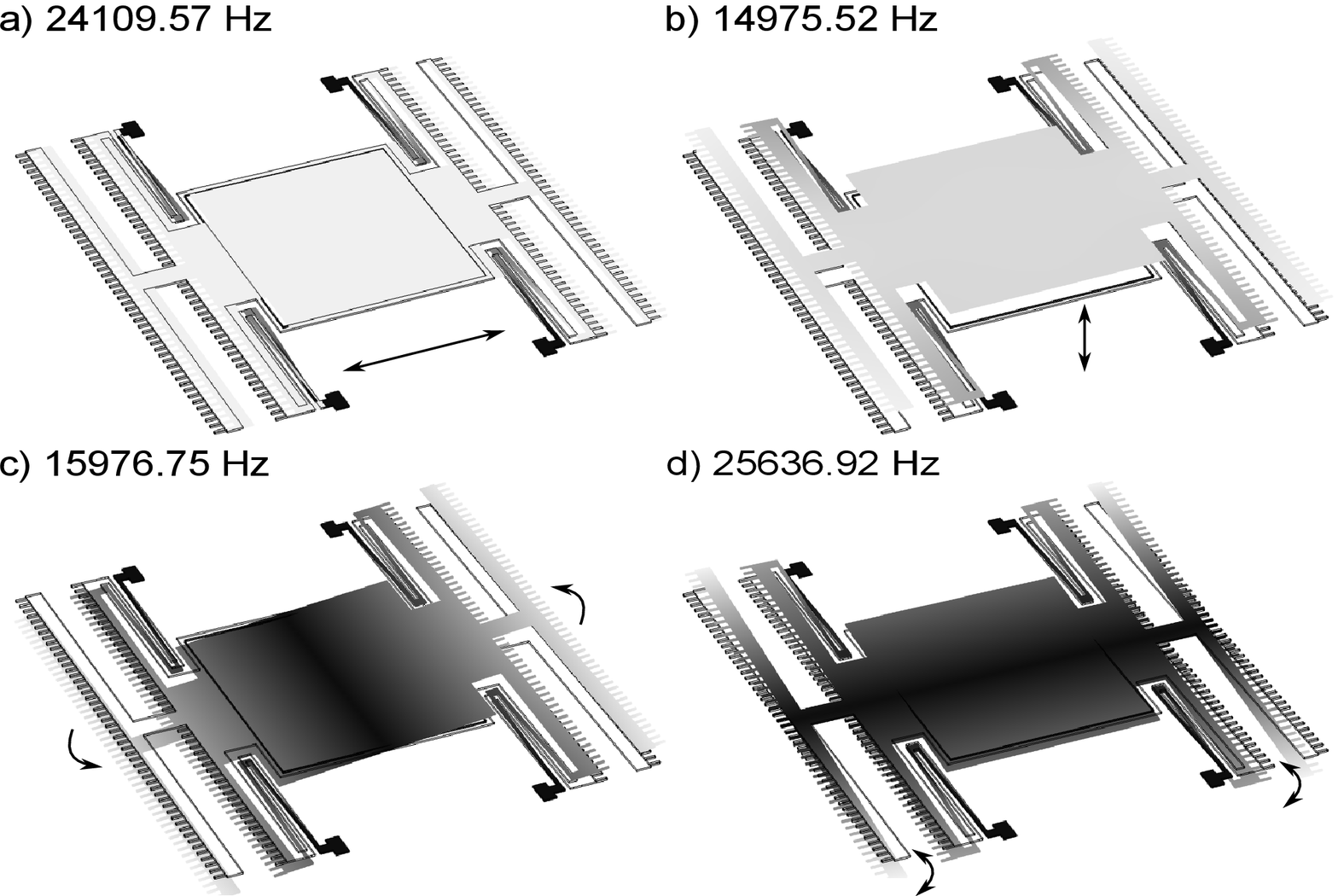}
\caption{\label{fig:H1modes}}{Vibrational modes and their corresponding frequencies for an H1 device. a) shear mode. b) Out-of-plane mode. c) pivot mode about y-axis. d) pivot mode about x-axis.}
\end{figure}
The top plate is electrostatically coupled to the fixed electrodes through comb-like structured terminals. This allows the device to be actuated in shear motion, parallel to the substrate. The two fixed electrodes on opposite side form a pair of series capacitors with the center plate as a common electrode, which allows us to implement a differential capacitance measurement by forming a bridge circuit with an external ratio transformer (TEGAM 1011A). A schematic diagram of the measurement scheme is provided in Fig.~\ref{fig:circuit}. A carrier signal $V_{hf}$, at $150$~kHz, is provided by the internal oscillator of a Signal Recovery 7124 lock-in amplifier. The ratio transformer splits this carrier signal into opposite phases that are connected respectively to the two fixed electrodes. Vibrations of the top plate are induced by another ac voltage at a much lower frequency. The low frequency excitation (Agilent 33220A), $V_{lf}$, is added to $V_{hf}$ through a homemade summing amplifier. The frequency of $V_{lf}$ is swept through the resonance. As the device goes through resonance the off-balance signal is picked up by an Amptek A250 charge sensitive amplifier. The amplification factor, $\alpha$, of the A250 is given by the inverse of the feedback loop's capacitance $c_{f}=1$~pF. The signal is later demodulated by two lock-in amplifiers connected in series. The first lock-in amplifier (Signal Recovery 7124) demodulates the carrier signal and the second lock-in (Signal Recovery 5210) detects the amplitude of the oscillations of the plate at the driving frequency ($V_{lf}$) or its harmonics. A dc bias voltage $V_{b}$ was applied to one of the side electrodes of a device through a 2~k$\Omega$ resistor. The total electrostatic force on the actuation side of the device can then be calculated and is given by
\begin{eqnarray}
\label{eq:force}
&&F_{e}(t)=\frac{n\tau\epsilon\epsilon_{0}}{2d_{f}}\left(V_{b}+V_{0}\sin{\omega t}\right)^{2} \\
&&=\frac{n\tau\epsilon\epsilon_{0}}{2d_{f}}\left[\left(V_{b}^{2}+\frac{V_{0}^{2}}{2}\right)+2V_{b}V_{0}\sin{\omega t}-\frac{V_{0}^{2}}{2}\cos{2\omega t}\right]. \nonumber
\end{eqnarray}
This force can be divided into three components: $F_{0}$, $F_{1}$, and $F_{2}$. The first term, $F_{0}\propto V_{b}^{2}+V_{0}^{2}/2$, is the constant term in parenthesis on the second line of Eq.~\ref{eq:force}. The second term, $F_{1}\propto 2V_{b}V_{0}\sin{\omega t}$, is in phase with the excitation and exists only when a DC bias is applied. The last term, $F_{2}\propto\frac{V_{0}^{2}}{2}\cos{2\omega t}$, is at twice the applied frequency. $F_{1}$ can be written as
\begin{equation}
\label{eq:transduction}
F_{1}(t)=\frac{n\tau\epsilon\epsilon_{0}}{d_{f}}V_{b}V_{0}\sin{\omega t}=\beta V_{ac}(t),
\end{equation}
where $\beta=n\tau\epsilon\epsilon_{0}V_{b}/d_{f}$ is often called the electro-mechanical transduction factor.

\subsection{Frequency Sweeps}
\begin{figure*}
{\includegraphics[width=0.9\linewidth]{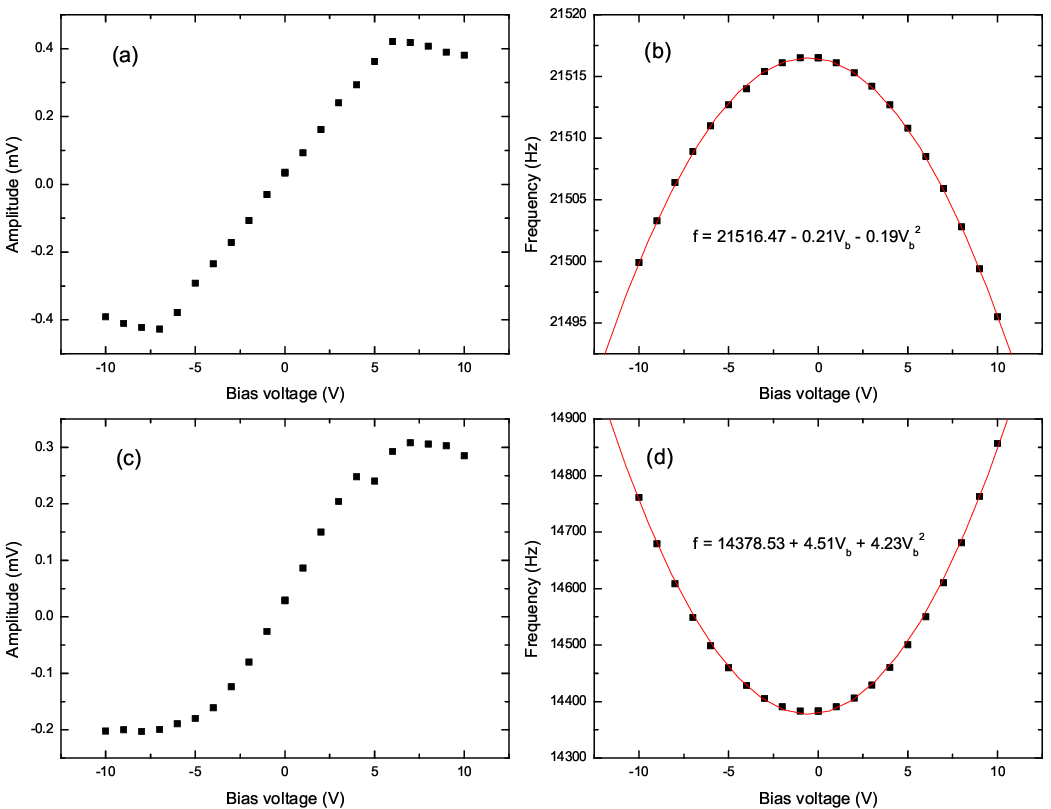}}
\caption{\label{fig:dcbias}}{Top: Amplitude (a) and resonance frequency (b) as a function of bias voltage for the shear mode of an H1 device. Bottom: Amplitude (c) and resonance frequency (d) as a function of bias voltage for the pivot mode of an H1 device. The lines represent a fit to a parabola. The resulting equations for the fit are displayed in each graph.}
\end{figure*}

After a device was wirebonded and the circuit was set up, the resonance peaks were found through frequency sweeps while keeping the device in a closed chamber under vacuum, at a pressure $P\sim 3$~mTorr. The amplitude and resonance frequency were determined from fitting to a Lorentzian curve. To connect the mechanical and the electrical systems, the transduction factor in Eq.~\ref{eq:transduction} is used to couple $F_{1}(t)$ and $V_{ac}(t)$ as well as $i(t)$ and the velocity of the movable plate, $v(t)$:
\begin{equation}
\label{eq:mecouple}
i(t)=\frac{d}{dt}[C(t)V_{b}]=V_{b}\frac{dC}{dx}\frac{dx}{dt}=\beta v(t).
\end{equation}
The previous expression also establishes a direct relation between the charge, $q(t)$, and the displacement, $x(t)$, through the transduction factor. A typical value for this factor is $\beta\approx 1.8\times 10^{-9} V_{b}$~C/m. Incorporating Eq.~\ref{eq:mecouple} into the solutions for the forced damped harmonic oscillator model \cite{landau-mech}, the in-phase ($q_{x}$) and out-of-phase ($q_{y}$) components of the charge detected can be found to be:
\begin{align}
\label{eq:peaks}
q_{x}(\omega)&=\frac{q_{0,x}(\omega_{0}^{2}-\omega^{2})\Delta\omega}{(\omega_{0}^{2}-\omega^{2})^{2}+(\omega\Delta\omega)^{2}} \\
q_{y}(\omega)&=\frac{q_{0,y}\omega\Delta\omega^{2}}{(\omega_{0}^{2}-\omega^{2})^{2}+(\omega\Delta\omega)^{2}}, \notag
\end{align}
where $q_{0,i}$ is a constant amplitude, $\omega_{0}$ is the natural frequency of the resonator, and $\Delta\omega$ is the half width at half maximum of the energy absorption peak. In all of our devices tested, two resonance peaks were found. These correspond to different modes of oscillations. One of the peaks was found at $\sim 14$~kHz and the other at $\sim 23$~kHz. The position of the peaks changes from device to device due to the variations in the fabrication process as explained above. In order to identify the type of mode corresponding for each resonance frequency, numerical simulations based on finite element analysis were performed using COMSOL Multiphysics. The results for the different modes and their corresponding frequencies for H1 are summarized in Fig.~\ref{fig:H1modes}. For these simulations, a fixed boundary condition (no motion) was used at the anchor point of the spring (see Fig.~\ref{fig:device}). Using the information from the simulations, the two modes found from the frequency sweeps were identified as the pivot mode about $y$-axis for the low frequency peak and the shear (in-plane) mode for the high frequency peak. The results of frequency sweeps, simulations, and analytical estimations of the resonance frequencies are summarized in Table~\ref{tab:table3} for the shear mode. Three different H1 devices were studied, indexed as H1, H1$_{t}$, and H1$_{m}$. The dimensions of these devices are the same, but the designs of the electrical connection from the bonding pads to the electrodes are different. For devices labelled with subscript ``t'', the connections were made ``thicker'', while for the ones labelled with subscript ``m'', the electrodes were covered with a gold layer to improve conductivity.

\subsection{DC Bias Test}

In the absence of a dc bias, the second term in Eq.~\ref{eq:force} ($F_{1}$) would disappear and no peak would be detected when attempting to detect a signal at the same frequency as the driving force. However, it was found through these studies that even at $V_{b}=0$ there was still a resonance peak. This means that there is an intrinsic bias voltage between the center electrode and the driving electrode. $V_{b}$ is consistent for many devices and conf. The detected resonance peak was found by scanning in ``$1f$'' mode in the second lock-in amplifier, which means the detected signal and the driving signal are at the same frequency. 

The bias voltage was changed from -10 to 10~V and the frequency was swept through resonance. The results for an H1 device with metallic electrodes (H1$_{m}$) are shown for the shear mode (top panel of Fig.~\ref{fig:dcbias}) and for the pivot mode (bottom panel of Fig.~\ref{fig:dcbias}). The amplitude increases linearly, as expected from Eq.~\ref{eq:force}, until it starts to saturate and to turn back down. This is a direct consequence of non-linearity starting to affect the peak due to high driving amplitudes. The resonance frequency follows a parabolic dependence on DC bias. The frequency of the shear mode decreases with the bias, while the frequency of the pivot mode is seen to increase with the bias.

\begin{figure}
\begin{center}
  \includegraphics[width=0.9\linewidth]{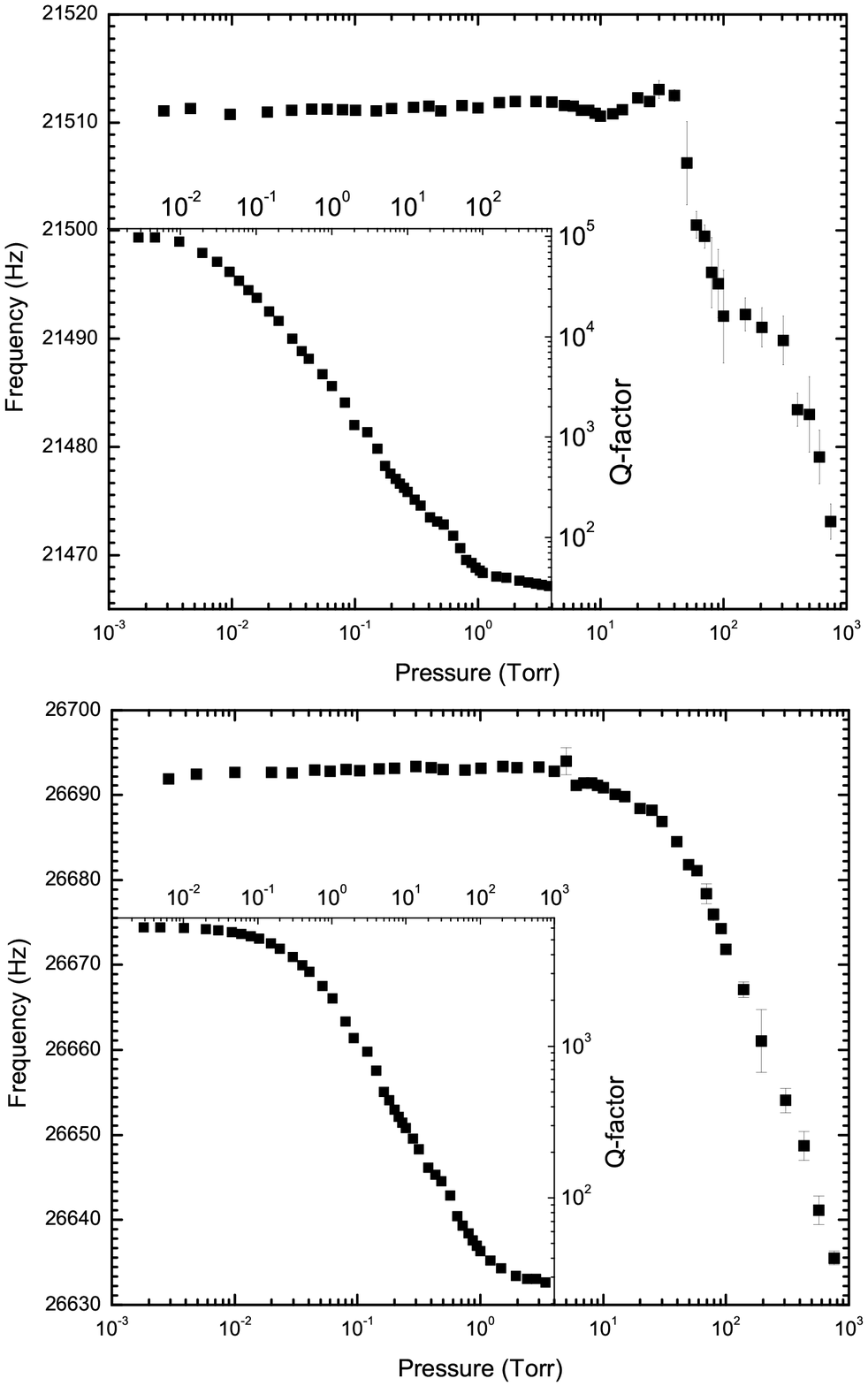}
\end{center}
\caption{\label{fig:Qpressure}}{Top: Q-factor and shear mode frequency of device H1$_{m}$ as a function of pressure. Bottom: Q-factor and shear mode frequency of device H2 as a function of pressure}
\end{figure}

\begin{figure}
\includegraphics[width=0.9\linewidth]{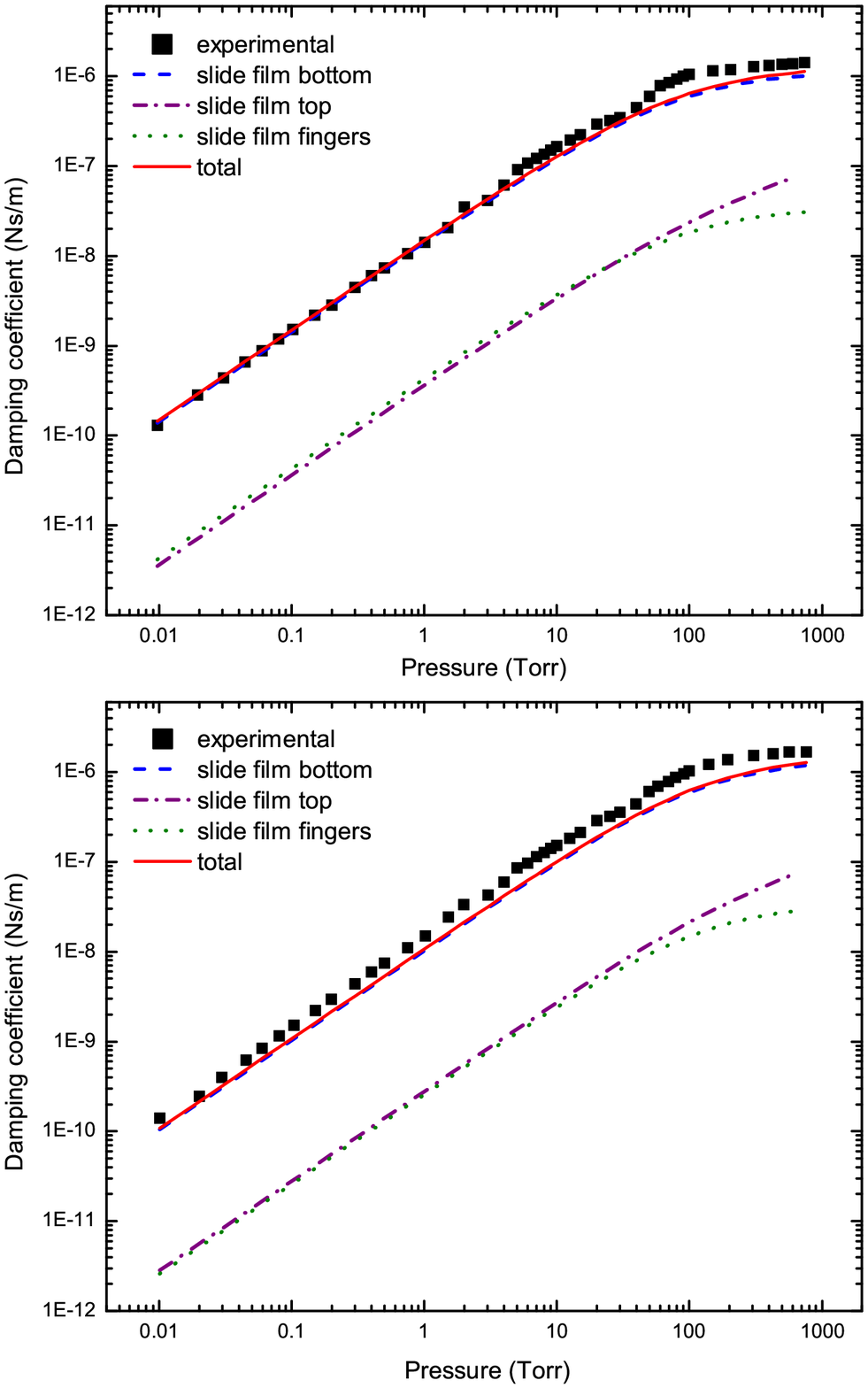}
\caption{\label{fig:damping}}{Top: Damping coefficient as a function of pressure for device H1. Bottom: Damping coefficient as a function of pressure for device H2. Calculated curves are shown for the damping due to the film (bottom), the top, and the fluid between the fingers.}
\end{figure}

\begin{figure}
\includegraphics[width=0.9\linewidth]{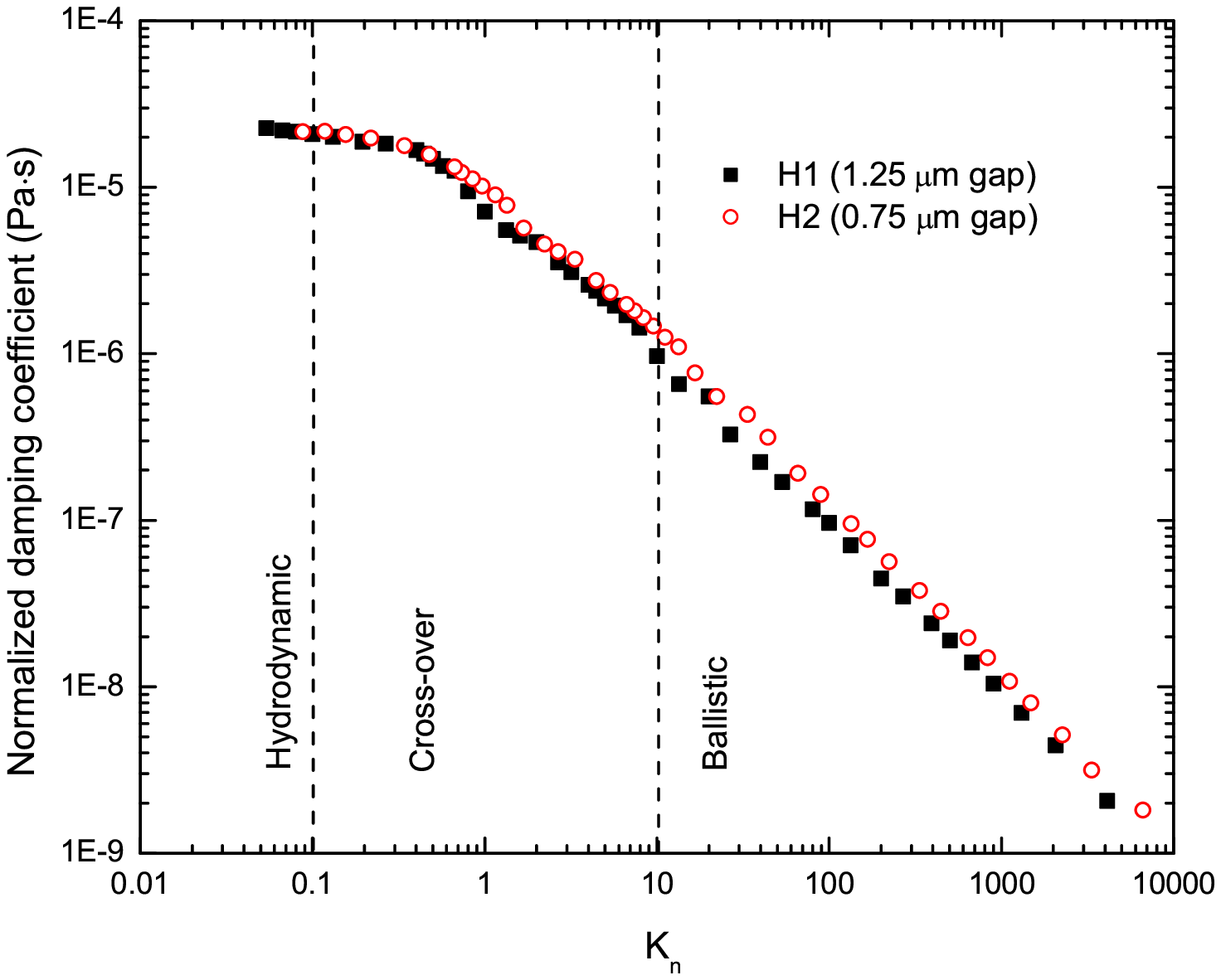}
\caption{\label{fig:normdamping}}{Damping coefficient as a function of pressure for both devices, H1 and H2, as a function of $K_{n}$. The vertical dashed lines separate the hydrodynamic, crossover and ballistic regimes.}
\end{figure}

Since the frequency of the $F_{2}$ component of the electrostatic force in Eq.~\ref{eq:force} is twice the excitation signal, the lock-in amplifier was set to detect vibrations at twice the input frequency (``$2f$'' mode) and the frequency was swept around half the value of the original position of the peak. Using the amplitude of the peak in ``$2f$'' mode, it is possible to obtain a measure of the intrinsic bias by taking a ratio with the amplitude in ``$1f$'' mode with no bias voltage applied. The ratio of the amplitudes for an H2 device is found to be 
\begin{equation}
\frac{A_{f}}{A_{2f}} = \frac{2V_{b}V_{0}}{V_{0}^{2}/2} = \frac{0.792~(mV)}{0.117~(mV)} = 6.76,
\end{equation}
from which we get $V_{b}=0.425~V$ using $V_{0}=0.25~V$. Similarly, using the shear mode peaks at ``$1f$'' and ``$2f$'', we found $V_{b}=0.482~V$. These values are in good agreement with the results obtained from the parabolic fit shown in Fig.~\ref{fig:dcbias}. Similar values are also obtained for all devices on different chips. 

\section{Room Temperature and Cryogenic Characterization}
\begin{figure*}
\includegraphics[width=0.9\textwidth]{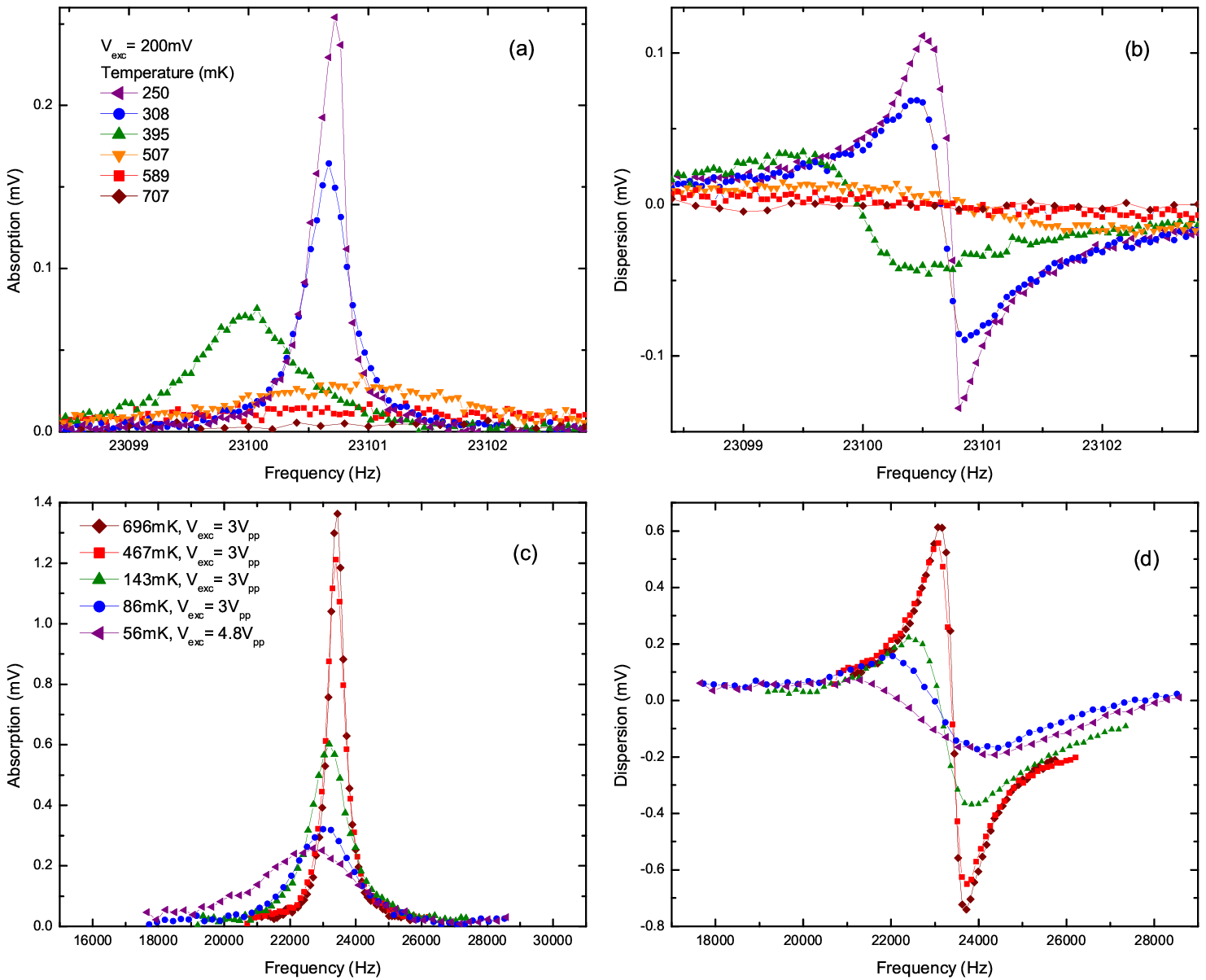}
\caption{\label{fig:hepeaks}}{Absorption (a) and dispersion (b) curves for an H1 device immersed in liquid $^{4}$He at different temperatures and at 2~bar pressure. Absorption (c) and dispersion (d) for an H2 device immersed in liquid $^{3}$He at different temperatures and at 3~bar pressure.}
\end{figure*}

\begin{figure*}
\includegraphics[width=0.9\textwidth]{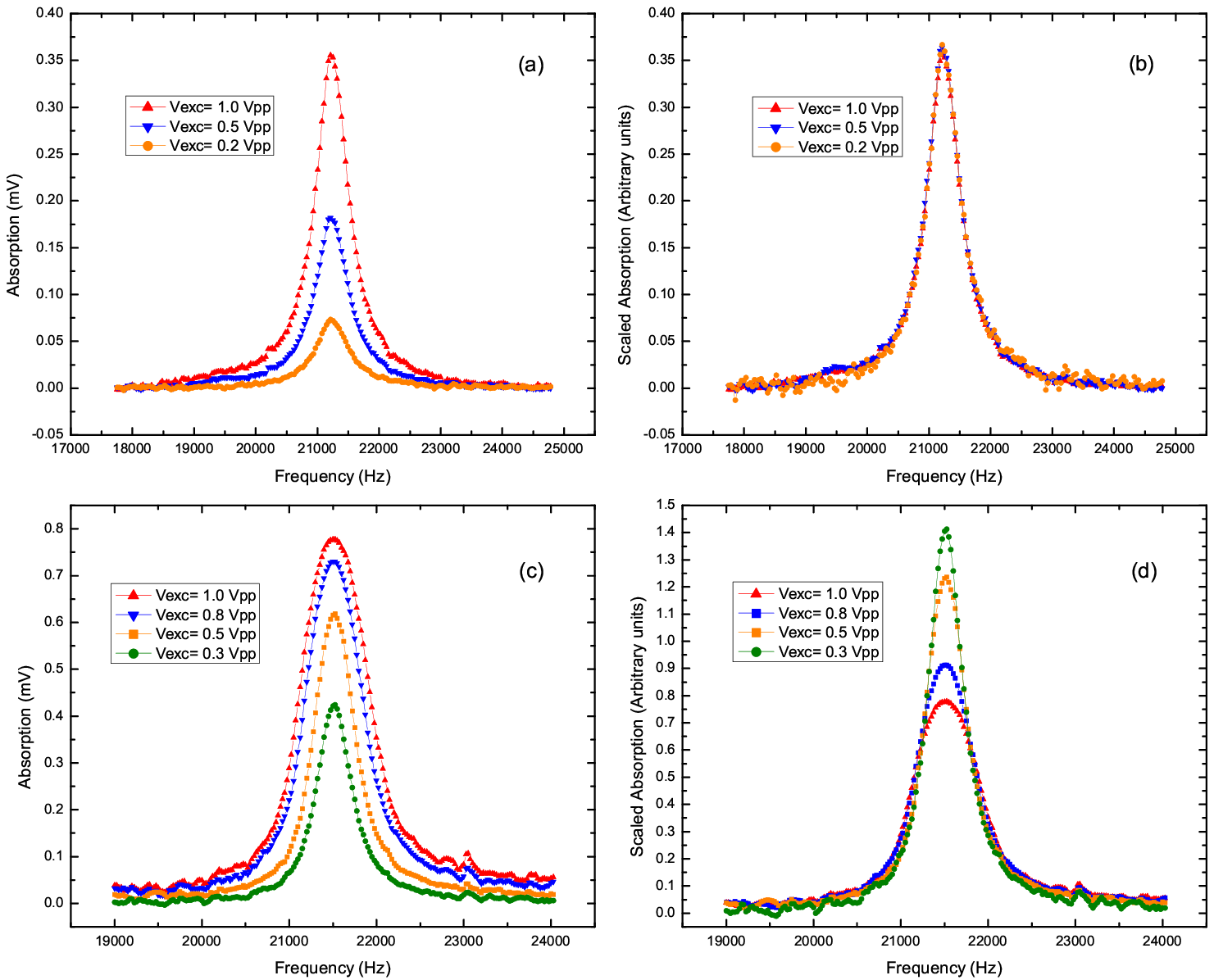}
\caption{\label{fig:sfpeaks}}{a) Absorption curves for an H1 device immersed in normal $^{3}$He at different excitations and T=276~mK. b) Absorption curves in normal $^{3}$He rescaled by the excitation voltage. c) Absorption curves for an H1 device immersed in superfluid $^{3}$He at different excitations and T$\approx$0.3~mK. d) Absorption curves in superfluid $^{3}$He rescaled by the excitation voltage.}
\end{figure*}

\subsection{Resonance as a Function of Pressure at Room Temperature}
The influence of air damping on the performance of miniature mechanical oscillators is a subject of utmost importance within the MEMS community from a technological standpoint. Additionally, due to the high surface to volume ratio, flow at the micro/nanoscale shows new phenomena not observed at the macroscopic scale \cite{cmho1998,klekinci2007,klekinci2008}. We carried out a rigorous study of the influence of air damping on our devices through a wide range of pressures, from 10~mTorr to 760~Torr.

Two different devices, H1 and H2, were studied individually while enclosed in a vacuum chamber, in which the pressure was controlled by venting and pumping the chamber through a needle valve from 10~mTorr to 1~atm. The frequency was swept through the shear mode resonance after the pressure was stabilized. The pivot mode was found to be significantly damped above 40~Torr due to the so-called squeeze film damping \cite{MBao2005}. Therefore, only the shear mode was studied. The frequency ($\omega_{0}$), width ($\Delta\omega$), and amplitude ($q_{0}$) of the peak were then obtained directly from the fit to the curves in Eq.~\ref{eq:peaks}. The quality factor, $Q$, was obtained by $Q=\omega_{0}/\Delta\omega$. The frequency and quality factor are plotted in Fig.~\ref{fig:Qpressure} for devices H1 (top) and H2 (bottom) as a function of pressure. The curve for $Q$ is essentially flat from 10 to 100~mTorr since the damping is mainly dominated by intrinsic dissipation. The quality factor then starts to decrease in an almost linear manner until  $P\sim 100$~Torr. At this point the quality factor begins to saturate again since in the hydrodynamic regime the damping becomes independent of pressure. However, the resonance frequency stays almost constant at pressures below $\sim 4$~Torr, and then it starts to decrease continuously as the pressure increases. This behaviour suggests a transition from a highly rarefied (ballistic) regime into a hydrodynamic regime. 

The coefficient of damping, $\gamma$, can be obtained experimentally from the width ($\Delta\omega$) as $\gamma=m\Delta\omega$. The intrinsic damping was obtained for both devices from the constant value of the width at low pressures. The experimentally determined damping coefficient as a function of pressure is shown in Fig.~\ref{fig:damping} for H1 (top) and H2 (bottom). In the two sets of data shown in Fig.~\ref{fig:damping} the intrinsic damping coefficient has been subtracted. The data therefore represent exclusively the damping coming from the fluid.

We followed the recipe given by Bruschi \emph{et al.} to estimate the theoretical value of the damping \cite{pbruschi2004}. The values of the damping coming from different parts, such as the center plate and the comb electrodes, can be calculated using their phenomenological expression for the damping of an oscillating plate in close proximity to a substrate, the so-called slide film damping:
\begin{equation}
\label{eq:dampingcoeff}
\gamma=\frac{A_{p}}{d}\left(\frac{\eta_{0}}{1+2K_{n}}\right),
\end{equation}
where $A_{p}$ is the area of the movable plate, $d$ is the gap between the plate and the substrate, $\eta_{0}$ is the viscosity at 1~atm, and $K_{n}$ is the Knudsen number, which is the ratio of the mean free path to the size of the gap. $K_{n}$ quantifies the rarefaction of the fluid, thus the term in the parentheses can be understood as an effective viscosity. To account for the finite size of the different structures of the device, an effective gap size was used. The effective gap is given by \cite{tveijola2001}
\begin{equation}
d_{eff}=\frac{d}{1+8.5d/l},
\end{equation}
where $l$ is the dimension of the oscillating structure along the motion of the plate. The slide film damping coming from the fingers (dotted line) and the fluid layer on top of the device (dash-dotted line) are almost two orders of magnitude smaller than the contribution from the fluid trapped between the gap (dashed line); see Fig.~\ref{fig:damping}. Thus, the total damping coefficient (solid line) is dominated almost exclusively by the damping coming from the fluid between the plates. These results are consistent with those obtained by Bruschi \emph{et al.} \cite{pbruschi2004}. Furthermore, when the damping coefficient is normalized by the geometrical constant $A_{p}/d$, both devices with different gaps exhibit a universal behaviour as a function of $K_{n}$, accross the various flow regimes, from hydrodynamic to ballistic (see Fig.~\ref{fig:normdamping}).

\subsection{Resonance as a Function of Temperature in Liquid Helium}
Low temperature studies in liquid $^{3}$He and in liquid $^{4}$He were conducted using a dilution refrigerator with a base temperature of 5~mK. The chip containing the MEMS device was housed in a low temperature experimental cell made of copper. The cell was loaded with the corresponding gas hypercritically to avoid damages from capillary condensation. The temperature was varied by applying heat to the mixing chamber of the dilution refrigerator. At each new heat setting the system was allowed to reach thermal equilibrium for a few hours. The temperature was measured using a calibrated ruthenium oxide thermometer located on the mixing chamber of the dilution unit and a melting pressure thermometer located on the nuclear demagnetization stage. Resonance properties were studied in both superfluid $^{4}$He and normal liquid $^{3}$He below 1~K. Resonances were obtained for the shear mode of an H1 device in $^{4}$He while H1 and H2 devices were used in $^{3}$He. All the measurements at low temperatures were conducted in ``$1f$'' mode using appropriate levels of dc bias.  This scheme is advantageous over the ``$2f$'' mode in terms of heat dissipation, which is a crucial aspect in low temperature experiments. The main source of Joule heating is from the electrical connections from the bonding pads and through the serpentine springs (see Fig.~\ref{fig:device}). Using the capacitance of the device and the calculated resistance of electrical connections from the bonding pad to the device, the Joule heating in this device is estimated to be $\approx$ 3~fW for 1~V$_{pp}$ ac excitation. We also have conducted measurements in superfluid $^{3}$He down to 300~$\mu$K and no appreciable heating was observed. 

The absorption and dispersion curves are shown in Fig.~\ref{fig:hepeaks} for superfluid $^{4}$He (top panels) between 250 and 707~mK and for normal liquid $^{3}$He (bottom panels) between 56 and 696~mK. A strong temperature dependence on the resonance was observed. In the case of $^{4}$He, the width of the resonance decreases as the temperature decreases. Below 1~K, the main mechanism of dissipation in superfluid $^{4}$He is through phonon excitations, which decrease as $~T^{4}$ in this temperature range. In contrast, the width of the peak in $^{3}$He increases as temperature decreases. Below its Fermi temperature (1~K), liquid $^{3}$He begins to display properties of a Fermi system. In the Fermi liquid regime, the viscosity of $^{3}$He increases very rapidly as $\sim 1/T^{2}$, which explains the broadening of the resonance as the temperature decreases. Further details of these experiments will be published elsewhere. Some preliminary results in superfluid $^{4}$He were discussed previously \cite{MGonzalez2011,Gonzalez2012-He4}.

Finally, we conducted a preliminary test of an H1 oscillator while submerged in superfluid $^{3}$He at 29~bar. In order to cool down below the base temperature of the dilution refrigerator (5.4~mK), adiabatic nuclear demagnetization was performed using an 8 Tesla superconducting magnet. After cooling into the superfluid regime, resonance peaks were obtained at the lowest temperature, which was determined to be $\approx$0.3~mK using a Pt NMR thermometer. The peaks obtained at this temperature are shown in Fig.~\ref{fig:sfpeaks}. As a function of increasing excitation voltage the peak is seen to become broader. This effect is more clearly seen when each measured peak amplitude is scaled by the excitation voltage. In normal $^{3}$He, the scaled resonance curves collapse into a single lorentzian line, indicating the resonance is in the linear regime (see Fig.~\ref{fig:sfpeaks}). However, in the case of superfluid $^{3}$He, the scaled curves do not collapse into a single lorentzian and instead deviate from the lorentzian shape and are reduced as a function of increasing excitation. This nonlinear behaviour could highlight a non-trivial dissipation mechanism in the liquid which is characteristic of the superfluid state.

\section{SUMMARY}
A new experimental tool based on MEMS oscillators was developed for use in low temperature physics experiments. This work provides a detailed roadmap of the development process from conception, design, and fabrication to application. Details on the experimental setup and the frequency response are presented. The quality of the surface was studied in detail by using AFM. Two length scales were identified from the autocorrelation function at $0.137$ and $0.591$~$\mu$m. From a height histogram the grain size was estimated to be $10.4$~nm.

The devices were studied as a function of pressure between 10~mTorr and 1~atm. This allowed us to study its resonance properties from a ballistic to a hydrodynamic regime. A model of slide film damping was found to be in excellent agreement with our experimental data.  We also demonstrate the use of these MEMS oscillators in a cryogenic environment and obtained resonance peaks of two devices immersed in liquid $^{4}$He and $^{3}$He below 1~K. The devices show great potential in a wide range of experiments at low temperatures. In particular, they might provide a powerful tool to study properties of liquid helium at the micro/nanoscale.

\section*{ACKNOWLEDGEMENTS}
We would like to acknowledge Hyoungjeen Jeen and Amlan Biswas at the University of Florida for facilitating the equipment and expertise necessary to carry out the AFM measurements. This work was supported by the National Science Foundation through DMR-0803516 and DMR-1205891 (YL).

\end{document}